\documentclass[letterpaper,twocolumn,showpacs,prl,aps,floatfix]{revtex4-1}

\usepackage{verbatim} 
\usepackage[latin1]{inputenc}
\usepackage{bm}
\usepackage{multirow,amssymb,amsbsy,amsmath}
\usepackage{graphicx}
\makeatletter
\usepackage{pifont}
\makeatother
\usepackage{color}  
\definecolor{gray}{rgb}{0.6,0.6,0.6}
\definecolor{Green}{rgb}{.0,.8,.5}

\def\vx{\boldsymbol{r}}

\begin{document}

\title{Photon-assisted confinement-induced resonances for ultracold atoms}

\author{Vicente Leyton$^1$}
\author{Maryam Roghani$^1$}
\author{Vittorio Peano$^2$}
\author{Michael Thorwart$^1$}

\affiliation{ 
  $^1$I.\ Institut f\"ur Theoretische Physik,
  Universit\"at Hamburg, Jungiusstra{\ss}e 9, 20355 Hamburg, Germany
  \\ 
  $^2$Institute for Theoretical Physics II,
  Friedrich-Alexander-Universit\"at Erlangen-N\"urnberg,
  Staudtstra{\ss}e 7, 91058 Erlangen, Germany }

\date{\today}

\begin{abstract}
We solve the two-particle s-wave scattering for an ultracold atom gas 
confined in a quasi-one-dimensional trapping potential which is
periodically modulated. The interaction between the atoms is included 
in terms of Fermi's pseudopotential.  For a modulated isotropic transverse
harmonic confinement, the atomic center of mass and relative degrees of
freedom decouple and an exact solution is possible.  We use the Floquet approach
to show that additional photon-assisted resonant scattering channels open up due
to the harmonic modulation. Applying the
Bethe-Peierls boundary condition, we obtain the general scattering
solution of the time-dependent Schr\"odinger equation which is universal at low
energies. The binding energies and the effective one-dimensional scattering
length can be controlled by the external driving.
\end{abstract}


\maketitle 

The ability to accurately control the effective atomic interactions in
ultracold atom gases has opened the doorway to novel exciting physics in the
recent years. The currently available experimental tools allow for a powerful
implementation of analog quantum simulators realized by cold-atom assemblies
\cite{Bloch12,Struck11}. In the regime of strong atomic interactions, the
quantum gas becomes scale-invariant and shows 
universal physical aspects quantified in terms of a few dimensionless
coefficients. For instance, the three-dimensional (3D) $s$-wave  scattering
length $a$ characterizes the interatomic interactions and has to be compared to
the mean interatomic distance  which is of the order of the particles'
 inverse momenta $k^{-1}$. It can be controlled over several
orders of magnitude via the help of a magnetic field tuned accross a Feshbach
resonance 
\cite{Feshbach99}. In 3D, the tunability of the scattering length, for instance,
reveals the cross-over from the weakly interacting Bardeen-Cooper-Schrieffer
superfluid state, where $1/(ka)\to -\infty$, to the strongly interacting
Bose-Einstein condensate of dimer molecules, where $1/(ka)\to
+\infty$ \cite{Bloch12}. 

In quasi-one-dimensional (1D) gases, another relevant length scale appears in
form of the transverse confinement length $a_\perp$. Then, the scattering of two
tightly confined quantum particles is known to induce universal
low-energy features in form of confinement-induced resonances (CIRs)
\cite{Olshani1,Olshani2,Olshani3}.  At low energies,  only the
transverse ground state of the confining potential is significantly
populated whereas the higher transverse states can be only virtually populated
during the elastic collisions. In this regime, the remaining scattering
processes in the longitudinal direction can be characterized by the effective 1D
interaction strength $g_{1D}$. It is governed by a single parameter, being the
ratio of $a$ and the zero-point-fluctuation length scale $a_\perp$,
irrespective of the details of the confinement. 

By tuning the confinement strength (or the $3D$ scattering length
via a Feshbach resonance \cite{Feshbach99}) across a CIR, it is possible to
cross over from strongly repulsive to strongly attractive interactions.  CIRs
have been observed in a strongly confined 1D gas of fermionic K atoms
\cite{Moritz05} and of bosonic Cs atoms \cite{Haller09}. This has allowed to
investigate the cross-over from a strongly repulsive Tonks-Girardeau gas to a
strongly attractive Super-Tonks-Girardeau gas \cite{Haller09}. CIRs have also
been observed in a strongly interacting 2D Fermi gas \cite{Frohlich11} and in
mixed dimensions \cite{LamporesiPRL2010}.

In close analogy to a Feshbach resonance, the CIR occurs when the
continuum threshold for the lowest transverse state (the {\em open channel\/})
has the same energy as a bound state formed by two particles being in some
transverse excited states (the {\em closed channels\/}) \cite{Olshani2}. Put
differently, the transverse orbital degrees of freedom of the confined atoms
play the same role as the internal atomic spin degrees of freedom for a Feshbach
resonance \cite{Feshbach99}.

When the transverse confinement of two equal atom species is purely harmonic,
 only one such bound state exists, leading to a single universal CIR
\cite{Olshani1,Olshani2,Olshani3}. This feature can be traced back to 
the separability of the center of mass and relative coordinates \cite{Peano}. In
turn, a multitude of CIRs appears when these degrees of freedom are no longer
separable, i.e., for a mixture of different species \cite{Peano}, anisotropic 
\cite{Drummond11,Sala12,Haller} 
 and anharmonic confinement
\cite{Peano,Kestner,Haller,Drummond11,Sala12,Sala13}, and in mixed dimensions
\cite{MassignanPRA2006,LamporesiPRL2010}. Dipolar CIRs
have been predicted to occur also in presence of long-range anisotropic
interactions between different atomic angular momentum states 
\cite{Schmelcher13}. Coupled CIRs have also been predicted at higher energies
when all partial scattering waves are taken into account \cite{Schmelcher13b}. 

As an alternative to the ``orbital'' Feshbach resonance to control the atomic 
scattering, a time-dependent modulation of
internal atomic states generates an optical Feshbach resonance
\cite{FedichevPRL}. It occurs when the optical radiation resonantly
couples two atoms in their respective electronic ground state to a molecular
state formed by electronically excited states, i.e., the relevant closed
channels in this case. This mechanism has been also experimentally demonstrated 
\cite{ZelevinskyPRL2006}. 

\begin{figure}[t]
\includegraphics[width=65mm]{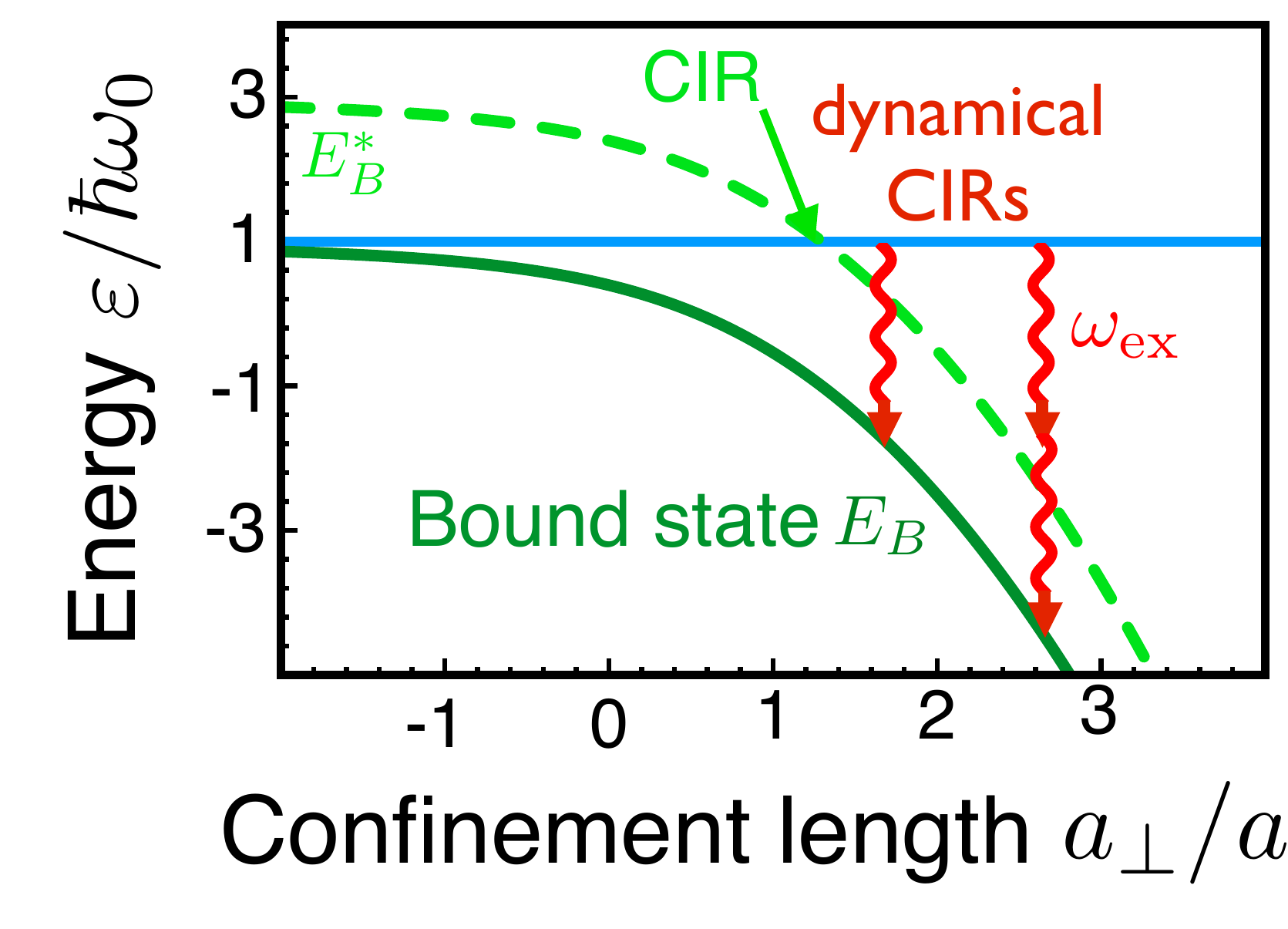}
\caption{\label{fig0} Sketch of the energy level scheme. A standard  CIR occurs
when 
the energy of the scattering atoms at the continuum threshold (horizontal line, 
blue online) matches the energy $E^*_B$ of a bound state formed by the closed
channels (dashed line, light green online). The energy $E_B$ of the real
bound state is shown as a continuum line (dark green online). A photon-assisted
CIR occurs when the binding energy $\hbar\omega_0-E_B$ matches the energy of
$m$ photons. The photon energy $\hbar\omega_{\rm ex}$ is indicated by wavy lines
at the photon-assisted CIRs  for the single-photon  and two-photon  resonances.
The quasienergy is the energy folded on an energy interval $\hbar\omega_{\rm
ex}$}
\end{figure}

In this work, we propose a novel mechanism to coherently manipulate the
scattering properties of cold atoms in quasi-1D traps via a time-dependent
RF modulation of the trapping potential. We consider a tight trap where
the atoms have been initially cooled to their transverse ground state. The trap
eigenfrequency $\omega_0$ is modulated by a periodic field such that 
$\omega^2(t)=\omega^2_0-F\cos \omega_{\rm ex}t$. The modulation is switched on
adiabatically in order to keep all  atoms in the same transverse Floquet 
quantum state. We show that a new type of CIR  is induced by the
virtual molecular recombination of the atoms (in their electronic ground state)
during the collision. This process is mediated by the emission of $m$ virtual 
photons  by the scattering atoms at the continuum threshold $\hbar\omega_0$, see
the level scheme in Fig.\ \ref{fig0}. Thus, the
virtual transition to the bound state with energy $E_B$ becomes resonant when 
$\hbar(\omega_0- m\omega_{\rm ex})=E_B$, leading to a series of photon-assisted 
CIRs.
This process is fundamentally different from any type of Feshbach resonances as
it does not involve a bound state formed by the {\em closed\/} channels. By
tuning the modulation field parameters, the photon-assisted CIRs can easily be
tuned.

In a frame of reference where the centre of mass of the two atoms is at rest,
the two-body problem can be mapped to the scattering of a trapped single
particle, with the reduced mass $\mu$, by a central potential
$U(\vx)$. The centre of mass and relative degrees of freedom decouple for a
harmonic trap also in presence of a parametric time-dependent modulation. 
The Hamiltonian can then be written as
\begin{equation} \label{eq:Ham2}
 H(\vx,t) = H_0(t) +U(\vx), \quad
  H_0(\vx,t) = \frac{p^2 }{2\mu}  
  + \frac{1}{2}\mu\, \omega^2(t) r^2_\perp.
\end{equation}
We set the $z$-direction as the direction of free
evolution. The confinement is defined over the $x-y$-plane,
implying that $\vx_{\perp}=(x,y)$. Furthermore, we choose the frequency
$\omega(t)$ such that the solutions of the classical equation of motion
$\ddot{x}+\omega^2(t)x=0$ are stable. We
consider the limit where the interaction range is much shorter than
the amplitude of zero point fluctuations $a_\perp =
(\mu \omega_0)^{-1/2}$ in the static trap ($\hbar=1$). Thus, the
scattering dynamics is governed by a single parameter, the 3D scattering length
$a$.  In this limit,
the low-energy interparticle interaction $U(\vx)$ can be described by
Fermi's pseudopotential \cite{Huang} %
\begin{equation}
 U({\vx}) = \frac{2\pi  a}{\mu} \delta ({\vx})
\frac{\partial}{\partial  r} r .
\end{equation}
We consider the atoms initially in the adiabatic
transverse ground state and having a longitudinal momentum $k$. They are
described by the incoming wave function 
\begin{equation}
\psi_{\rm in}(t)=\exp [-i\varepsilon t]
\exp [ikz] u_0(x,t)u_0(y,t),
\end{equation}
where $\varepsilon$ is the  quasienergy $\varepsilon=k^2/2\mu+\nu$ of the atoms.
Here,  the time-periodic functions $u_n(x,t)$ are the Floquet eigenstates of the
parametrically driven harmonic oscillator with quasienergy $(n+1/2)\nu$
\cite{Brown91,Grifoni,SM}. Notice that, when the driving is switched off
adiabatically, $\nu\to\omega_0$ and the $u_n(x)$ become the eigenstates of the
harmonic oscillator. 

\begin{figure}[t!]
\includegraphics[width=65mm]{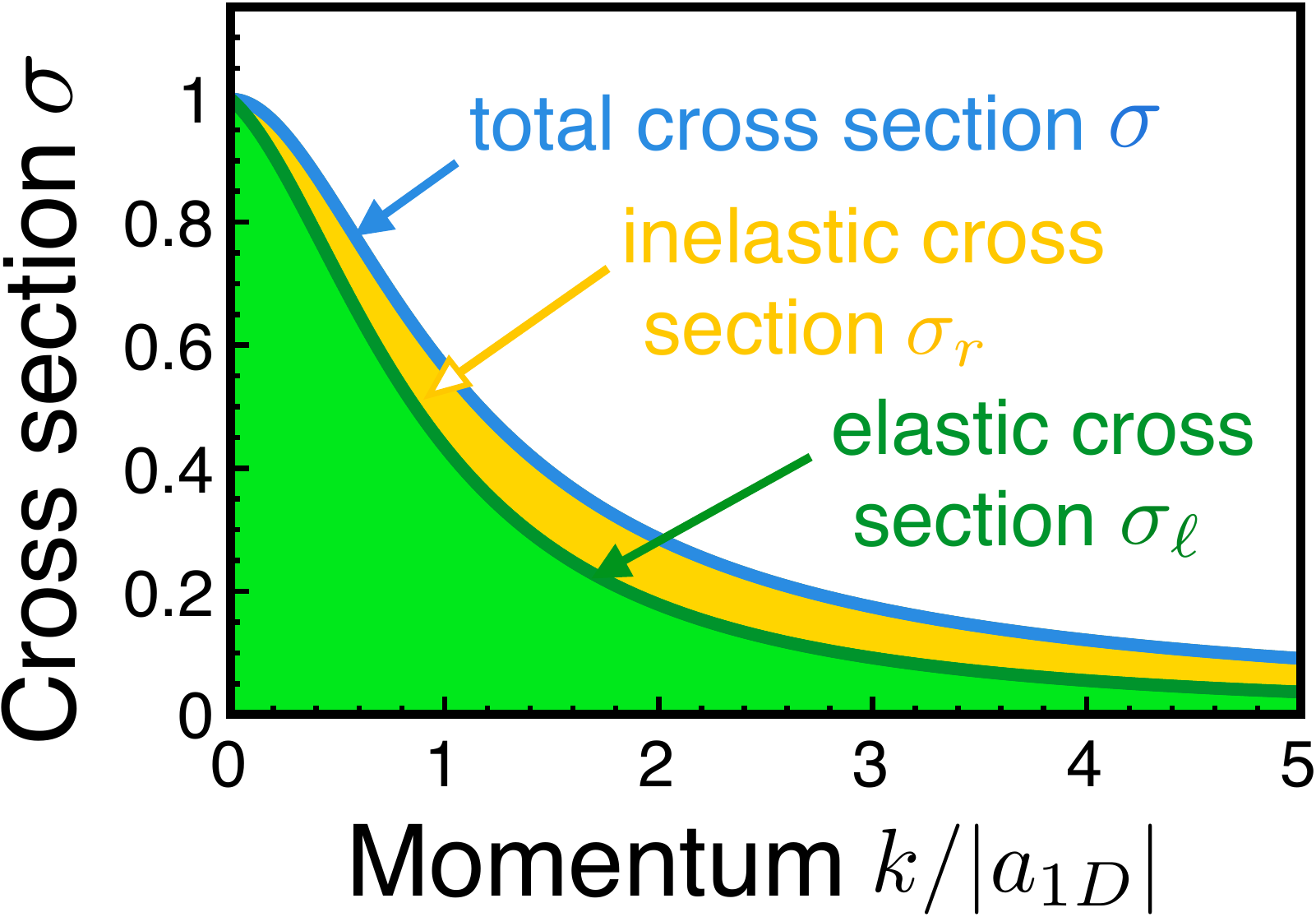}
\caption{\label{fig2} Effective scattering cross sections in quasi-1D: total
($\sigma$) and
elastic ($\sigma_\ell$) cross section as a function of the relative longitudinal
momentum $k$ for $|{\rm Im }\, a_{\rm 1D}|/|a_{\rm 1D}|=0.15$.}
\end{figure}

Our goal is to compute the full solution $\psi(t)$ which includes the scattered
wave $\psi_{\rm out}(t)$ such that $\psi(t)=\psi_{\rm in}(t)+\psi_{\rm out}(t)$.
It is most convenient to introduce the time-periodic Floquet state
$\phi(t)=\exp[i\varepsilon t]\psi(t)$ as the solution of the  eigenvalue problem
${\cal H}(\vx,t)\, \phi(\vx,t) = \varepsilon \,\phi (\vx,t)$, where 
${\cal H}\equiv H - i\partial_t$ is the Floquet Hamiltonian
\cite{Grifoni}. With $T=2\pi/\omega_{\rm ex}$ being the external modulation
period, we can formally write 
\begin{equation}
  \label{eq:sol}
  \phi (\vx,t) = \phi_{\rm in}(\vx,t) +  \int_0^T \frac{dt'}{T}
  {\cal G}_\varepsilon (\vx,0;t,t') \frac{f (t')}{2 \mu},
\end{equation}
where $\phi_{\rm in}(t)=\exp [i\varepsilon t]\psi_{\rm in}(t)$ and the second
term on the r.h.s.\ represents the outgoing scattered
wave function $\phi_{\rm out}(t)$. The intergral kernel ${\cal G}_\varepsilon =
({\cal H}_0 - \varepsilon - i0)^{-1}$ is the
retarded Floquet-Green's function with ${\cal H}_0 = H_0 -
i \partial_t$. This wave function has to fullfill the Bethe-Peierls
boundary condition
\begin{equation}
  \label{eq:sctt}
\phi(\vx\to 0,t) \simeq\left(1-\frac{r}{a}\right) \frac{f(t)}{4\pi r}\, ,
\end{equation}
with the Bethe-Peierls amplitude $f(t)$ yet to be determined. 
For large $z$, the asymptotic scattered wave can be decomposed into partial
waves as
\begin{equation}\label{eq:asout}
 \phi_{\rm out}({\bf{r}})\approx\sum_{\substack{\mathbf{n} \\ 
m= {\rm open}}}
S^m_{\mathbf{n}}\left(\frac{k}{k_{nm}}\right)^{1/2}e^{ik_{nm}|z|}
u_{n_x}(x,t)u_{n_y}(y,t)\,.
\end{equation}
Here, $|S^m_{\mathbf{n}}|^2$ is the probability of the atoms to be excited into 
the transverse state with quantum number ${\bf n}=(n_x,n_y)$ after absorbing
$m>0$ photons from the field and thereby acquiring the momentum $k_{nm}$, which
is determined via 
\begin{equation}\label{eq:enconsv}
\frac{k_{nm}^2}{2\mu}+(n+1)\nu=\varepsilon+m\omega_{\rm ex},\qquad n=n_x+n_y
\, .
\end{equation} 
The number of available open channels depends on $m$.  The
S-matrix elements $S^m_{\mathbf{n}}$ are determined by the Bethe-Peierls
boundary condition Eq.\ \eqref{eq:sctt} (for technical detials, see
Ref.\ \cite{SM}) as 
\begin{equation}\label{eq:reflrate}
S^m_{\mathbf{n}} =\frac{i}{2\sqrt{kk_{nm}}T}\int_0^{T}\!\! dt^\prime
e^{im\omega_{\rm ex}t}u^*_{n_x}(0,t')u^*_{n_y}(0,t')f(t')\,.
\end{equation} 
Next, we insert Eq.\ \eqref{eq:sol} into
Eq.\ \eqref{eq:sctt} and define a scalar product $\langle
f|g\rangle=T^{-1}\int_0^T dt \, f^*(t)g(t)$ on the Hilbert space of
time-periodic functions. Then, we can derive an inhomogeneous linear equation 
 for the Bethe-Peierls amplitude $f(t)$ as 
\begin{equation}\label{eq:bound2}
 \left(\zeta_\varepsilon+\frac{a_\perp}{a}\right)  |f \rangle  =-4\pi {\cal
N}^{-1/2}|{\rm in}\rangle 
 \ .
\end{equation} 
Here, we have introduced the regularized integral kernel
\begin{equation}
  \label{eq:regker}
  \langle t|\zeta_\varepsilon |t'\rangle =  
  \frac{2\pi a_\perp}{ \mu}   
  \left[ {\cal G}_\varepsilon(\vx,0;t,t') -
    \delta(t-t')\frac{T\mu}{2\pi r}
    \right]_{\vx \rightarrow 0} 
\end{equation}
and the normalized vector $|{\rm in}\rangle$ via $\langle t|{\rm
in}\rangle={\cal N}^{1/2}a_\perp\phi_{\rm in}(0,t)$.
For  small initial momenta, $k\ll k_{nm}$, the scattering is dominated by
elastic collisions. In this regime, it is convenient to divide the kernel
$\zeta_\varepsilon$ into a smooth part $\tilde\zeta_\varepsilon$, 
 that can be evaluated for $k=0$, and the contribution of the channel of
the incoming atoms, where the $k$-dependence is retained. This yields 
\cite{SM}
\begin{equation}\label{eq:foft}
 |f\rangle = -4\pi {\cal N}^{-1/2}\frac{ ka_{\rm 1D}}{ ka_{\rm 1D}-i}\left( 
\tilde{\zeta}_\nu + \frac{a_\perp}{ a}\right)^{-1}|{\rm in}\rangle \, .
\end{equation}
Here, it is convenient to introduce the effective 1D 
scattering length $a_{1D}$ for the longitudinal scattering according to 
\begin{eqnarray}\label{eq:a1D}
 \frac{a_\perp}{a_{\rm 1D}} &=& -2\pi{\cal N}^{-1} \langle {\rm in}| \left[
\tilde{\zeta}_\nu
      +\frac{a_\perp}{a} \right]^{-1}|{\rm in} \rangle \, .
\end{eqnarray}
Since the operator $\tilde{\zeta}_\nu$ is not Hermitian, it acquires an
imaginary part whose meaning is discussed further below. 
Using Eq.\ \eqref{eq:foft} in Eq.\ \eqref{eq:reflrate}, we obtain the
$S$-matrix element 
\begin{equation}\label{S0-matrix}
S^0_{00}=-\frac{i}{k a_{\rm 1D}-i}\,.
\end{equation}
This is the probability amplitude for the reflection of a 1D  particle due to 
the effective scattering potential $U_{\rm 1D}=g_{\rm 1D}\delta(z)$ with complex
interaction strength $g_{1D}=-1/(\mu a_{\rm 1D})$. Its imaginary part refers 
to the loss of atoms in the excited transverse states which occurs due to
inelastic scattering processes into other channels provided by the modulation. 
From this, we obtain the elastic cross section, which is the probability of an
elastic scattering event, as $\sigma_\ell=|S^0_{00}|^2$ and its inelastic
counterpart as $\sigma_r=1-\sigma_\ell-|1+S^0_{00}|^2$,
respectively.  From Eq.\ \eqref{S0-matrix}, we find 
\begin{eqnarray}\label{eq:cross}
&&\sigma_\ell=\left( 1+k^2|a_{\rm 1D}|^2-2k\, {\rm Im}\, a_{\rm
1D}\right)^{-1},\nonumber \\
&&\sigma_r=-2\sigma_\ell k\, {\rm Im}\, a_{\rm 1D},
\end{eqnarray}
(${\rm Im}\, a_{\rm 1D}<0$). The effective quasi-1D scattering cross
sections  (for the case
$|{\rm Im }\, a_{\rm 1D}|/|a_{\rm 1D}|=0.15$) are shown in  Fig.\ \ref{fig2}. 
The probability of an elastic scattering event tends to one for small realtive
momenta of the scattering atoms,
 $k\ll1/|a_{\rm 1D}|$. On the other hand, the scattering is dominated by 
inelastic scattering events for comparatively larger momenta, $k\gg|{\rm Im}\,
a_{\rm 1D}|^{-1}$. Hence, when $|a_{\rm 1D}|$ becomes smaller than the typical
longitudinal de Broglie wavelength of the atoms a scattering resonance results.

Formally, the 1D scattering length
$a_{\rm 1D}$ can be expressed in terms of the spectrum $\{\lambda_m\}$ and
the right and left eigenvalues $|v_m^R\rangle$ and  $|v_m^L\rangle$ of the
kernel $\tilde{\zeta}_\varepsilon$,
\begin{equation}\label{eq:a1D2}
\frac{a_\perp}{ a_{\rm 1D}} =-2\pi{\cal N}^{-1}\sum_m\frac{\langle {\rm
in}|v^R_m\rangle\langle v^L_m|{\rm in}\rangle}{\lambda_m+a_\perp/ a} \, .
\end{equation}
Hence, if several eigenvectors  significantly overlap with the vector $|{\rm
in}\rangle$ of the incoming wave, more than one scattering
resonances may occur. The resonances in $|a_\perp/a_{\rm 1D}|$ are well resolved
Lorentzian peaks with their center determined by the resonance condition 
$a_\perp/a={\rm Re}\, \lambda_m$ and with their line width 
given by ${\rm Im} \, \lambda_m$, if their mutual distance exceeds
the corresponding widths. 
\begin{figure}[t!]
\includegraphics[width=75mm]{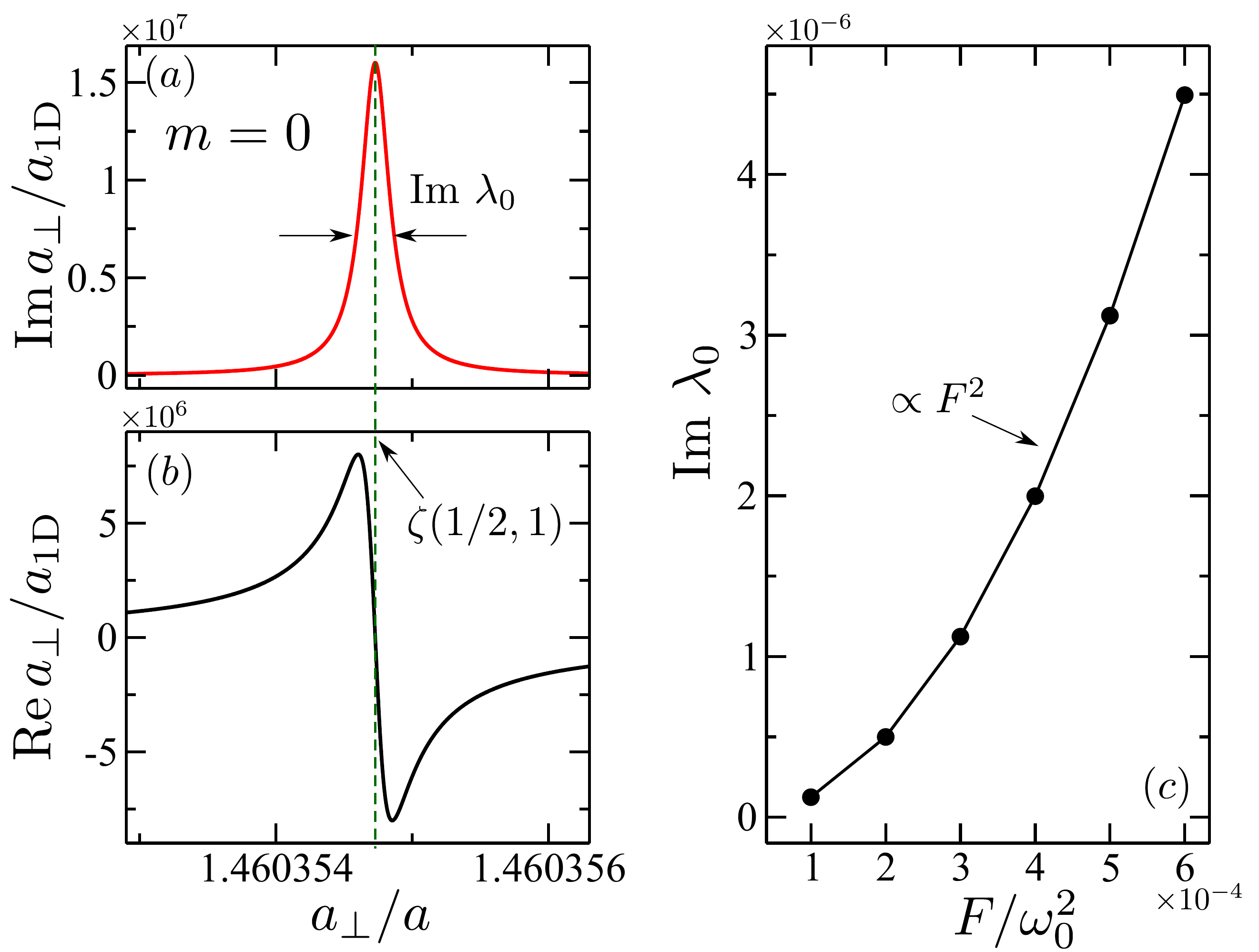}
\caption{\label{fig3}Photon-assisted broadening of the standard CIR. The
imaginary and real parts of
$a_\perp/a_{\rm 1D}$ are shown in (a) and   (b), respectively,  for $\omega_{\rm
ex}=1.2 \omega_0$ and $F=10^{-4}\omega_0^2$. The dashed vertical line indicates 
the resonance position in absence of driving,
$a_\perp/a=-\zeta(1/2,1)$. The resonance is broadened by  driving-induced
inelastic
transitions to  excited transverse channels. (c) The width of the CIR has a   
quadratic dependence on the driving amplitude,
Im $\lambda_0 \propto F^2$.}
\end{figure}

The kernel $\tilde\zeta_\varepsilon$ can be computed fully  analytically only
for the trivial case
$F=0$, see Supplemental Information \cite{SM}. In this case,  the left and 
right eigenvectors, $|v_m^L\rangle$ and
$|v_m^R\rangle$, are plane waves, 
$\langle t|v_m^R\rangle^*=\langle v^L_R|t\rangle=\exp[im\omega_{\rm ex}t]$, and
the incoming wave is given by $|{\rm in}\rangle=|v^R_0\rangle$.  Thus, the
overlap is
zero for $m\neq 0$ and we recover the standard result: there is only one CIR  
for $a_\perp/a=-\lambda_0=-\zeta(1/2,1)$
($\zeta(x,y)$ is the Hurwitz zeta function) when the energy of the
virtual bound state formed by the
closed channels, coincides with the continuum threshold \cite{Olshani1}. 

 If the driving is weak and far away from any parametric resonance
\begin{equation}
F\ll \omega_{\rm ex}^2,\qquad |2\omega_0-m\omega_d|\gg \omega_d (4F/\omega_d^2)^m,
\end{equation}
we expect the eigenvalues  $\{\lambda_m\}$ of the kernel $\zeta_\varepsilon$ to only smoothly 
deviate from their values for $F=0$. To obtain a specific result,  we evaluate
and diagonalize the kernel $\tilde\zeta_\varepsilon$ with a semi-analytical
procedure outlined in the Supplementary Material \cite{SM}.  The standard CIR
for a weak non-resonant driving is
shown in Figs.\ \ref{fig3} (a) and (b).  The zero-photon resonance is clearly
broadened by the inelastic collisions. In Fig.\ \ref{fig3} (c) we show that  the
width ${\rm Im}\lambda_0\propto F^2$.
%
\begin{figure}[t!]
\includegraphics[width=60mm]{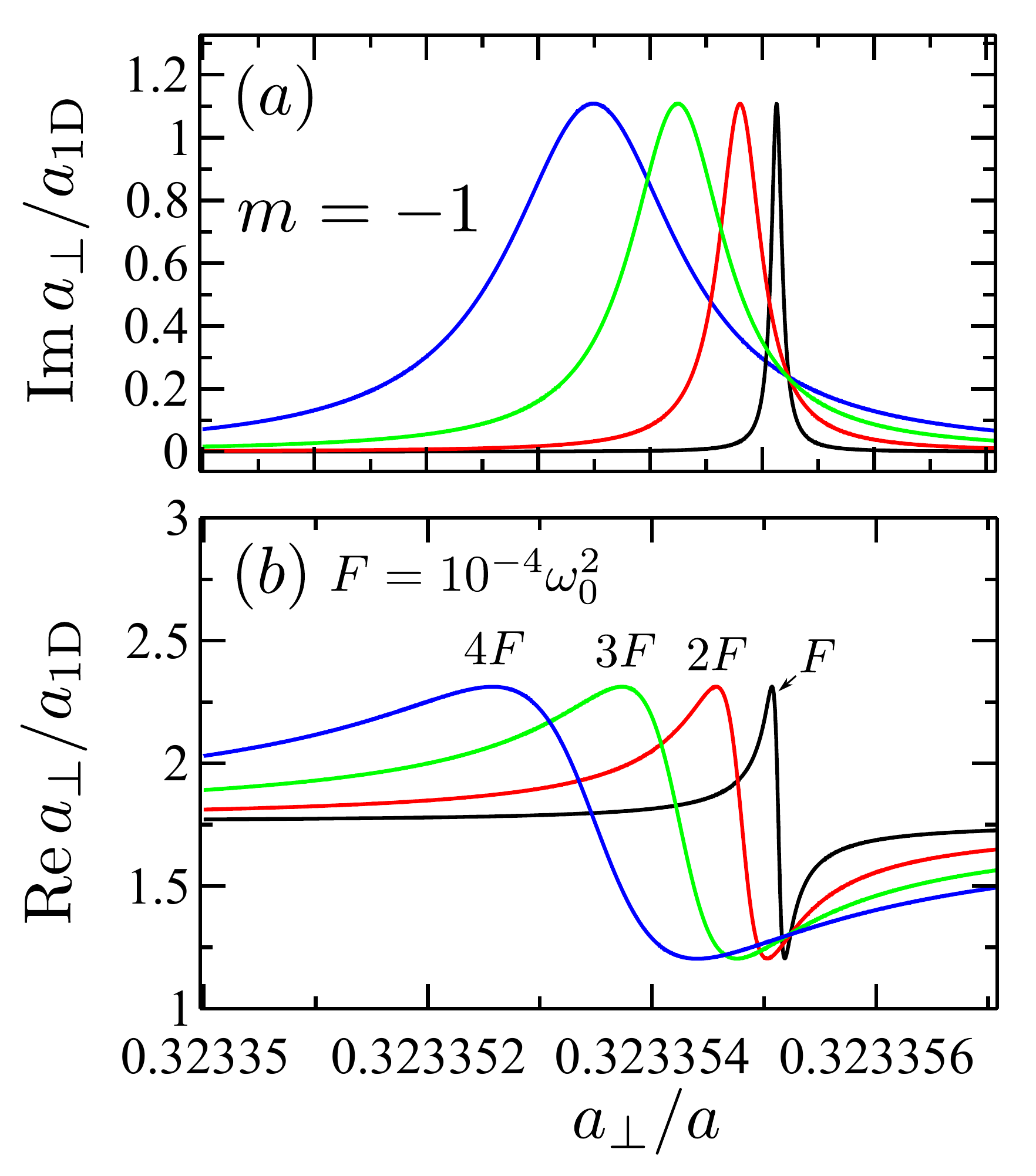}
\caption{\label{fig4}Photon-assisted confinement-induced resonances (imaginary
part of
$a_\perp/a_{\rm 1D}$ in (a) and real part in (b)) due to a one-photon
absorption from and subsequent emission into the driving field ($m=-1$) for
$\omega_{\rm ex}=1.2 \omega_0$ and for varying modulation amplitudes $F$.}
\end{figure}

 For a  finite driving  there is no selection rule preventing the  remaining
eigenvectors of the kernel $\tilde{\zeta}_\varepsilon$ to yield a finite
contribution in the r.h.s of Eq.\ \eqref{eq:a1D2} to $a_{1D}$. For weak
non-resonant driving, we can label the eigenvalues $\lambda_m$ with the number
$m$ of radiofrequency photons which are virtually absorbed (emitted for $m<0$)
and later re-emitted (re-absorbed)  during an elastic  collision. From Eq.\
\eqref{eq:a1D2}, we see that the contribution to $a_{1D}$ from the processes
where $m$ photons are virtually absorbed  is largest for $a_\perp/a=-{\rm
Re}\lambda_{m}$. In the Supplemental material, we show that, for the case $m>0$,
this occurs when the $m$-photon transition from the continuum 
threshold $\hbar\omega_0$ to the virtual bound state with energy $E_B^*(m)$ formed by
the transverse channels which are still closed after the absorption of
$m$ photons [with transverse energy $E>E_M={\rm Int} [m\omega_{\rm
ex}+\omega_0]$] is resonant, $E^*_B (m)=\omega_0+m\omega_{\rm ex}$.  We do not
expect that these processes lead to a scattering resonance because the
corresponding bound states leak very quickly into the open channels. In fact,
the imaginary part ${\rm Im}\lambda_m$ can be shown to be finite even for $F\to
0$, ${\rm Im}\lambda_m\gtrsim 1$ \cite{SM}. On the other hand,the contribution
to $a_{1D}$ from processes where the photons are first emitted ($m<0$) is
largest for \cite{SM}
\begin{equation}
\frac{a_\perp}{a}=-{\rm Re}\lambda_{m}\approx-\zeta(1/2, |m|\omega_{\rm ex}/2\omega_0)
\end{equation} 
Notice that the energy $E_B$ of the molecular bound state (for $F=0$) is given
by $a_\perp/a=-\zeta(1/2, (E_B-1)/2\omega_0)$ \cite{Olshani2}. Hence, the 
processes where $|m|$-photons are virtually emitted  leads to the largest
enhancement of scattering  when  the molecular recombination
accompanied by the emission of $m$ photons is resonant, $\omega_0-
|m|\omega_{\rm ex}\approx E_B$.  Since  $\lim_{F\to 0}{\rm Im} \lambda_m=0$
\cite{SM} (the molecular bound state can only dissociate because of the
driving), these processes lead to  sharp CIRs. The resonance in the case of
resonant emission of a single photon ($m=-1$) is shown in Fig.
\ref{fig4} for different values of $F$. Notice that, the   scattering resonances
investigated here involve  the true molecular bound state and not   a virtual
bound state formed by the closed channels. Therefore, they are not 
Feshbach-type resonances but rather the dynamical equivalent of a shape
resonance (a scattering resonance that occurs when a
generic potential has a bound state closed to the continuum threshold)
\cite{Landau}.

{\it Conclusions - } We have shown that the s-wave scattering of two atoms
confined in a tight quasi-1D trap can be coherently controlled by a RF 
modulation of the transverse confinement. The
scattering in the atom cloud in the adiabatic transverse ground
state can be efficiently described as atoms in a 1D waveguide interacting via a
short range interaction. In this case, the coupling constant $g_{1D}$ acquires
an imaginary part and incorporates inelastic scattering into the transverse 
excited states. This mechanism generates a 
new kind of photon-assisted or dynamical confinement-induced resonances.   The
results are universally valid in the regime of low energies. The
dynamical confinement-induced resonances are another example in which
photon-assisted processes carry signatures of atomic interparticle interactions
which have also been seen in the photon-assisted tunneling in a Bose-Einstein
condensate \cite{Eckardt05a}. Such processes could be used for
avoided-level-crossing spectroscoy in strongly interacting quantum gases
\cite{Eckardt08}.  

{\it Acknowledgements - } We acknowledge support from the DFG
Sonderforschungsbereich 925 ``Light induced dynamics and control of
correlated quantum systems'' (project C8).

\end{document}